\documentclass[twocolumn, 10pt]{article}

\usepackage[square,numbers,sort&compress,comma]{natbib}

\usepackage[utf8]{inputenc}
\usepackage{graphicx}
\usepackage{tabularx}
\usepackage{amsmath}
\usepackage{amssymb}
\usepackage{times}
\usepackage{caption}

\usepackage[pagewise]{lineno}
\usepackage[version=4]{mhchem}
\usepackage{siunitx}
\topmargin - 12pt 
\renewenvironment{abstract}%
              {
               \small
               {\bfseries \abstractname}
               \par
               \vspace{10pt}
              }

\renewcommand\abstractname{Abstract}

\newcommand{\nomenclature}
              [1]
              {
               \bgroup
               \flushleft
               \small\bf
               #1
               \par
               \egroup
              }

\renewcommand{\section}
              [1]
              {
               \bgroup
               \flushleft
               \small\bf
               \refstepcounter{section}
               \arabic{section}. #1
               \par
               \egroup
              }

\renewcommand{\subsection}
              [1]
              {
               \bgroup
               \flushleft
               \small\em
               \refstepcounter{subsection}
               \arabic{section}.
               \arabic{subsection}. #1
               \par
               \egroup
              }

\renewcommand{\subsubsection}
              [1]
              {
               \bgroup
               \flushleft
               \small\em
               \refstepcounter{subsubsection}
               \arabic{section}.
               \arabic{subsection}.
               \arabic{subsubsection}. #1
               \par
               \egroup
              }

  \newcommand{\acknowledgement}
              [1]
              {
               \bgroup
               \flushleft
               \small\bf
               #1
               \par
               \egroup
              }

  \newcommand{\sectionbib}
              [1]
              {
               \bgroup
               \flushleft
               \small\bf
               #1
               \par
               \egroup
              }

\begin{document}

\title{\LARGE Numerical Analysis of the Stability of Iron Dust Bunsen Flames}
\author{ {\large T. Hazenberg$^{a,*}$, D. Braig$^{a}$, J. Mich$^{a}$, A. Scholtissek$^{a}$, C. Hasse$^{a}$}\\[10pt]
         {\footnotesize \em $^{a}$ Technical University of Darmstadt, Simulation of reactive Thermo-Fluid Systems, Germany}
       }

\date{}

\small
\baselineskip 10pt

\twocolumn[\begin{@twocolumnfalse}
\vspace{50pt}
\maketitle
\rule{\textwidth}{0.5pt}

\begin{abstract}
  This article presents numerical simulations of the response of an iron dust Bunsen flame to particle seeding changes.
  A validated numerical model is used to study the impact of particle seeding fluctuations on flame stability.
  Simulations are conducted for the Bunsen setup in the right-side up and up-side down configuration.
  No significant differences in flame response are identified in flame stability between the right-side up and up-side down configurations.
  We find that the Bunsen flame is surprisingly robust to abrupt changes in particle loading.
  The sudden change in particle loading does not excite any intrinsic instabilities in the flame.
  Based on our results, the iron dust flames are robust to imposed fluctuations.
  We hypothesize that this is due to the lack of a feedback mechanism between the burned temperature and the heat release rate.
  This mechanism is present in conventional, chemistry-driven, gaseous flames.
  However, such a mechanism is absent in iron dust flames because the combustion of individual iron particles is limited by oxygen diffusion, which is insensitive to temperature.
\end{abstract}

\vspace{10pt}
\parbox{1.0\textwidth}{\footnotesize {\em Keywords:} metal fuels; Bunsen flame; iron dust combustion; stability;}
\parbox{1.0\textwidth}{\footnotesize {\em Preprint submitted for review}}
\rule{\textwidth}{0.5pt}
\vspace{10pt}

\end{@twocolumnfalse}] 

\section{Introduction}
In recent years, metal fuels have gained significant interest as an alternative renewable energy carrier, e.g.,~\cite{2007_Mignard_reviewspongeiron,2015_Bergthorson_DirectCombustionRecyclable,2018_Bergthorson_RecyclableMetalFuels}.
Combustion of fine powders of metals air releases heat and produces metal oxide powder.
When the metal oxide powder is captured after combustion, it can be reduced using renewable energy.
As metals have a high energy density, they might be a good alternative, e.g., to hydrogen, especially when transportation of the energy carrier is required.
The combustion of metal does not involve carbon, and initial studies indicate that \ce{NO_{x}} emissions are also minimal~\cite{2024_Ravi_Nitrogenoxideformation,2024_Prasidha_Towardsefficientmetal,2025_Duebal_chemicalreactornetwork}.
In other words, metals have the potential to be carbon-free, minimal-emission, and renewable energy carriers.

A thorough understanding of iron combustion is required to develop reliable, efficient, low-emission iron dust combustors.
To develop reliable, efficient and low-emission iron dust combustors a thorough understanding of iron dust combustion is required.
In recent years, the knowledge of the community of single particle combustion has significantly increased, e.g.,~\cite{2015_Soo_ReactionParticleSuspension,2022_Ning_Temperaturephasetransitions,2024_Mich_Modelingoxidationiron,2023_Fujinawa_Combustionbehaviorsingle,2023_Panahi_Combustionbehaviorsingle,2024_Nguyen_Ignitionkineticlimited}.
In several of these studies, satisfactory agreement has been obtained between experimental observations and numerical results.
Parallel to developing single particle experiments and models, the community has developed experiments and models to improve our understanding of flame propagation in iron dust clouds, e.g.,~\cite{2017_Soo_ThermalStructureBurning,2021_Hazenberg_StructuresBurningVelocities,2019_McRae_StabilizedFlatIron,2023_Fedoryk_ExperimentalInvestigationLaminar,2024_Hulsbos_HeatFluxMethod,2023_Gool_ParticleEquilibriumComposition,2023_Mich_ComparisonMechanisticModels,2023_Ravi_Effectparticlesize,2024_Krenn_Evaluationnovelmeasurement}.
However, unlike the single particle work, minimal agreement has been obtained between numerical and experimental results.
To our knowledge, only in the very recent article of Hazenberg et al.~\cite{2025_Hazenberg_AnalyzingIronDust} is an excellent agreement between model and experiment shown.

One of the major challenges in validating iron dust models is the limited results of well-controlled, properly understood experiments for estimating the burning velocity.
The most significant difficulty in setting up reliable experiments for measuring the burning velocity is maintaining a stable flame.
Particle seeding is often said to be the culprit of this; providing a stable and homogenous stream of particles is challenging.
This article aims to provide an exploratory assessment of particle seeding-induced flame instabilities.
To this end, we will perform simulations with our validated model~\cite{2025_Hazenberg_AnalyzingIronDust} on the experimental setup of Fedoryk et al.~\cite{2023_Fedoryk_ExperimentalInvestigationLaminar}.
As a first step, fluctuations in particle loading are modeled via a single abrupt (and homogenous) increase in particle loading.
The flame's response to a sudden change in particle loading is then recorded and analyzed.

This article is divided into the following parts:
First, the numerical model and simulation parameters are presented.
Second, the results of the simulations are presented and analyzed.
Finally, conclusions, a discussion, and an outlook are provided for the numerical results.

\section{Numerical methods}
In our previous publication~\cite{2025_Hazenberg_AnalyzingIronDust}, we presented and validated the numerical framework we will utilize in this work.
Therefore, we will limit the discussion of the numerical framework to providing the governing equation and refer to our previous work for more details.
After giving the governing equations, we will present the relevant boundary conditions. 
As simulations are conducted with the same particle size distribution and on the same experimental setup as in the previous publication, we not provide much detail on these.

\subsection{Gas phase model}
The gas phase is modeled via the usual conservation equations of mass, momentum, enthalpy and species:
\begin{equation}
  \frac{\partial}{\partial t} \rho 
+ \frac{\partial}{\partial x_i}\left(\rho u_{\mathrm{g},i}\right) 
= 
  S_{\mathrm{prt}, m} \mathrm{,}
\end{equation}
\begin{align}
    \frac{\partial}{\partial t}\left(\rho u_{\mathrm{g},i}\right) 
& + \frac{\partial}{\partial x_j}\left(\rho u_{\mathrm{g},i} u_{\mathrm{g},j}\right) \\
& =
  - \frac{\partial}{\partial x_i}p 
  + \frac{\partial}{\partial x_j}\tau_{i j} 
  + \rho g_i+S_{\mathrm{prt}, u_i} \mathrm{,} \nonumber
\end{align}
\begin{align}
    \frac{\partial}{\partial t}(\rho h)
& + \frac{\partial}{\partial x_j}\left(\rho u_{\mathrm{g},j} h\right) \\
& = 
    \rho u_{\mathrm{g},j} g_j
  - \frac{\partial }{\partial x_j}q_j
  + S_{\mathrm{prt}, h} \nonumber
\end{align}
and
\begin{align}
    \frac{\partial}{\partial t}\left(\rho Y_k\right)
& + \frac{\partial}{\partial x_j}\left(\rho u_{\mathrm{g},j} Y_k\right) \\
& = 
    \frac{\partial}{\partial x_j}\left(\rho D_k \frac{\partial }{\partial x_j}Y_k\right) 
  + S_{\mathrm{prt}, Y_k} \mathrm{,} \nonumber
\end{align}
where $\rho$ is the density, 
$u_{\mathrm{g},i}$ is the flow velocity in spatial dimension $x_i$, 
$p$ is the pressure, $\tau_{ij}$ is the stress tensor, 
$g_i$ is the gravity, $h$ is the enthalpy, $q_i$ is the heat flux, 
$Y_k$ is the mass fraction and $D_k$ is the diffusivity of species k. 
Finally, $S_{\mathrm{prt}}$ represents exchange terms between the continuous gas and the dispersed solid phase in the above equations.

\subsection{Particle model}
Particles are tracked via Newton's equations of motion, i.e.,
\begin{equation}
  \frac{\mathrm{d} x_{\mathrm{p},i}}{\mathrm{d} t} = u_{\mathrm{p},i}
\end{equation}
and
\begin{equation}
  m_{\mathrm{p}} \frac{\mathrm{d} u_{\mathrm{p},i}}{\mathrm{d} t} = m_{\mathrm{p}} g_i + F_{\mathrm{d}, i} + \left(u_{\mathrm{g},i} - u_{\mathrm{p},i}\right) \frac{\mathrm{d} m_{\mathrm{p}}}{\mathrm{d} t} \mathrm{,}
  \label{Eq:velocity_evolution}
\end{equation}
with $x_{\mathrm{p},i}$ the particle position and $u_{\mathrm{p},i}$ the particle velocity.
In the second equation, the first term is due to gravity, with $g_i$ the gravity vector, and the second term is the drag force provided by
\begin{equation}
  F_{\mathrm{d}, i}=\frac{1}{2} \rho_{\mathrm{g}} \left|u_{\mathrm{g}, i}-u_{\mathrm{p}, i}\right| \left(u_{\mathrm{g}, i}-u_{\mathrm{p}, i}\right) C_{\mathrm{d}} A_{\mathrm{p},\mathrm{c}} \mathrm{,}
\end{equation}
with
\begin{equation}
  C_{\mathrm{d}}=\frac{24}{Re_{\mathrm{p}, i}}\left(1+\frac{1}{6} Re_{\mathrm{p}, i}^{2 / 3}\right) \mathrm{,}
\end{equation}
where $A_{\mathrm{p},\mathrm{c}}$ is the cross-sectional area and $Re_{\mathrm{p}}$ the particle Reynolds number.
Finally, the last term in Eq.~\ref{Eq:velocity_evolution} is due to the change of particle momentum due to the change in particle mass, which, in this form, is only valid for a particle whose mass is increasing.

The evolution of the particle mass is tracked via
\begin{equation}
  \frac{\mathrm{d} m_{\mathrm{p}}}{\mathrm{d} t} = \dot{m}_{\mathrm{ox}} \mathrm{,}
  \label{Eq:mass_evolution}
\end{equation}
where $\dot{m}_{\mathrm{ox}}$ is the oxidation rate of the particle.
We can find the oxidation rate by assuming that the surface reaction rate equals the oxygen diffusion rate to the particle surface.
More specifically, the oxidation rate is found by solving the following algebraic equation for $Y_{\ce{O2},\mathrm{s}}$ (the mass fraction of gas phase oxygen at the particle surface),
\begin{align}
  \rho_{\mathrm{s}} & Y_{\ce{O2},\mathrm{s}} A_{\mathrm{p}} k_{\infty} \exp \left(\frac{-T_{\mathrm{a}}}{T_{\mathrm{p}}}\right) \label{Eq:Diffusion_reaction_balance} \\
& = - A_{\mathrm{p}} \frac{Sh \left(\rho D_{\ce{O2}}\right)_{\mathrm{f}}}{d_{\mathrm{p}}} \ln \left( 1 + B_{\mathrm{m}} \right) \mathrm{,} \nonumber
\end{align}
where $B_{\mathrm{m}}$ is the mass Spalding number given by
\begin{equation}
  B_{\mathrm{m}} = \frac{Y_{\ce{O2},\mathrm{s}} - Y_{\ce{O2},\mathrm{g}}}{1-Y_{\ce{O2},\mathrm{s}}} \mathrm{.}
\end{equation}
In the above equations, $\rho_{\mathrm{s}}$ is the gas density at the particle surface, $A_{\mathrm{p}}$ is the particle surface area, $k_{\infty}$ is the pre-exponential factor of the surface reaction, $T_{\mathrm{a}}$ is the activation temperature of the surface reaction, $Sh$ is the Sherwood number, $(\rho D_{\ce{O2}})_{\mathrm{f}}$ is the oxygen diffusivity evaluated inside the film layer, $d_{\mathrm{p}}$ is the particle diameter and $Y_{\ce{O2},\mathrm{g}}$ is the oxygen mass fraction in the bulk gas phase.
Eq.~\ref{Eq:Diffusion_reaction_balance} is solved by approximating the left and right-hand sides as a second-order polynomial in $Y_{\ce{O2},\mathrm{s}}$ and then identifying the roots of the resulting equation.
When the solution for $Y_{\ce{O2},\mathrm{s}}$ is known, one can use the left or right-hand side to compute $\dot{m}_{\mathrm{ox}}$.
Particle burnout is determined when there is no more iron in the particle.
To this end, the mass of iron is tracked inside the particle
\begin{equation}
  \frac{\mathrm{d} m_{\ce{Fe},\mathrm{p}}}{\mathrm{d}} = \frac{1}{s} \dot{m}_{\mathrm{ox}} \mathrm{,}
\end{equation}
where $s$ is the stoichiometric ratio of iron to oxygen.
We assume oxidation to \ce{FeO}, such that $s=55.85/15.99$.

Finally, the evolution of the particle temperature is tracked via
\begin{equation}
  m_{\mathrm{p}} c_{\mathrm{p},\mathrm{p}} \frac{\mathrm{d} T_{\mathrm{p}} }{\mathrm{d} t} = \dot{q}_{\mathrm{conv}} + \dot{q}_{\mathrm{rad}} + \dot{q}_{\mathrm{reac}} + \dot{m}_{\ce{O2}} h_{\ce{O2},s},
  \label{Eq:enthalpy_evolution}
\end{equation}
where $c_{\mathrm{p},\mathrm{p}}$ is the particle heat capacity, including the heat of phase changes, and the right-hand side includes convective heat transport, radiative heat loss, the heat of reaction, and energy transfer due to enthalpy transport.
The convective heat transfer includes a correction for the Stefan flow and is provided by
\begin{equation}
  \dot{q}_{\mathrm{conv}} = - A_{\mathrm{p}} \frac{Nu \lambda_{\mathrm{f}}}{d_{\mathrm{p}}}\left(T_{\mathrm{p}}-T_{\mathrm{g}}\right) \frac{\ln \left(1+B_{\mathrm{t}}\right)}{B_{\mathrm{t}}},
\end{equation}
the radiative heat loss is provided by
\begin{equation}
  \dot{q}_{\mathrm{rad}} = -\epsilon \sigma A_{\mathrm{p}} \left(T_{\mathrm{p}}^4-T_{\mathrm{g}}^4\right)\mathrm{,}
\end{equation}
and the heat release of reaction by
\begin{equation}
  \dot{q}_{\mathrm{reac}} = - h_{\ce{Fe}} \frac{\mathrm{d} m_{\mathrm{p},\ce{Fe}}}{\mathrm{d} t} - h_{\ce{FeO}} \frac{\mathrm{d} m_{\mathrm{p},\ce{FeO}}}{\mathrm{d} t} \mathrm{,}
\end{equation}
where $B_{\mathrm{t}}$ is the energy Spalding number, $Nu$ is the Nusselt number, $T_{\mathrm{g}}$ is the gas phase temperature, $\epsilon$ is the emissivity of the particle, and $\sigma$ is the Stefan-Boltzmann constant.
The model for the boundary layer assumes that the Lewis number is unity and the particle is stationary, such that $B_{\mathrm{t}}=B_{\mathrm{m}}$ and $Nu=Sh=2$. 
Finally, the particle thermodynamics (phase densities, heat capacities, etc.) are described with correlations from the NIST database~\cite{1998_Chase_NISTJANAFThemochemical}, including transitions between phases.

Based on Eq.~\ref{Eq:velocity_evolution}, Eq.~\ref{Eq:mass_evolution} and Eq.~\ref{Eq:enthalpy_evolution}, the source terms from the particle phase can be defined for a numerical cell as
\begin{equation}
  S_{\mathrm{prt}, m} = -\frac{1}{V_{\mathrm{cell}}} \sum_m \left(\frac{ \mathrm{d} m_{\mathrm{p}}}{ \mathrm{d} t}\right)_j n_j \mathrm{,}
\end{equation}
\begin{equation}
  S_{\mathrm{prt}, u_i} = -\frac{1}{V_{\mathrm{cell}}} \sum_m \left(F_{\mathrm{d},i}\right)_j n_j \mathrm{,}
\end{equation}
\begin{equation}
  S_{\mathrm{prt}, h} = -\frac{1}{V_{\mathrm{cell}}} \sum_m \left( \dot{q}_{\mathrm{conv}} + h_{\ce{O2}, \mathrm{s}}\frac{\mathrm{d} m_{\mathrm{p}}}{\mathrm{d} t}\right)_j n_j
\end{equation}
and
\begin{equation}
  S_{\mathrm{prt}, Y_{\ce{O2}}} = -\frac{1}{V_{\mathrm{cell}}} \sum_m \left(\frac{\mathrm{d} m_{\mathrm{p}}}{\mathrm{d} t}\right)_j n_j \mathrm{,}
\end{equation}
where $V_{\mathrm{cell}}$ is the volume of the corresponding cell, $m$ represents the number of parcels inside a cell, subscript $j$ represents the $j$th particle, and $n$ the number of particles per parcel.

\begin{figure}
  \centering
  \includegraphics{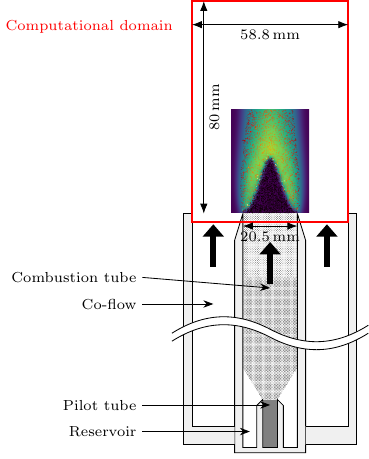}
  \caption{Schematic of the experimental setup~\cite{2023_Fedoryk_ExperimentalInvestigationLaminar}. 
  The computational domain is marked by a red box. 
  For illustrative purposes a sudden increase of the particle loading is drawn in the combustion tube.}
  \label{fig:expSetup}
\end{figure}
\subsection{Numerical Setup}
The simulations are conducted on the same numerical domain as in our previous work~\cite{2025_Hazenberg_AnalyzingIronDust}; a summary is provided here for completeness.
Fig.~\ref{fig:expSetup} provides an overview of the experimental setup by Fedoryk et al.~\cite{2023_Fedoryk_ExperimentalInvestigationLaminar}.
The combustion pipe is located in the center of the setup; particle seeding takes place in this pipe.
The length of this combustion pipe is chosen so the flow has time to develop and particles that this flow cannot lift fall.
While this cannot avoid in particle loading and flow conditions fluctuations, they are minimized.
As a result, an iron dust Bunsen flame can be stabilized above the combustion pipe - a co-flow of air shields the Bunsen flame to improve its stability further.

\begin{table*}
  \centering
  \caption{Boundary conditions specified for the computational domain.}
  \begin{tabular}{llll}
    \hline
    Surface & T & U & p \\ \hline
    Fuel & fixed value & fixed value  & fixed value  \\ 
    Co-flow	  & fixed value & fixed value & zero gradient  \\ 
    Wall  & fixed value & no slip & zero gradient  \\ 
    Surroundings  & zero gradient & inlet/outlet & wave transmissive  \\ 
  \end{tabular}
  \label{tab:BC}
\end{table*}
The simulations only consider the region inside the red box above the combustion pipe.
The area above the combustion pipe and co-flow is discretized with \num{1.1} million cells.
In Table~\ref{tab:BC}, an overview of the relevant boundary conditions is provided.

It is assumed that the seeding in the combustion pipe is ideal, i.e., besides statistical fluctuations, it is homogenous and constant in time.
A step change in the particle loading is utilized to model hypothetical fluctuations in particle loading.
This change in particle loading is homogenous over the inlet and does not exhibit local changes.
At the inlet, particles are injected, and their size is sampled from a log-hyperbolic~\cite{1977_BarndorffNielsen_Exponentiallydecreasingdistributions} distribution function, which has been fit to the experimental distribution function.
Each particle is injected at the terminal velocity with respect to the gas; if this results in a negative velocity, the particle is removed.

Finally, parcels are used instead of particles to reduce the computational load.
The smallest parcels (\SI{0.5}{\micro\meter}) contain around \num{2000} particles, and any parcel larger than \SI{6}{\micro\meter} is modeled as a single particle.
In between \num{0.5} and \SI{6}{\micro\meter}, the number of particles per parcel is chosen such that the total initial mass of the particles inside a particle is equal to that of a \SI{6}{\micro\meter} particle.
As OpenFOAM\copyright samples parcels from the particle size distribution, the log-hyperbolic distribution is scaled by the inverse of the particles per parcels function.
Scaling the particle size distribution in this way ensures that the particle size distribution matches that of the experiments and not the parcel size distribution.

\section{Results}
Two simulations are conducted to study the stability of iron dust flames.
Both simulations are performed at a mean inlet velocity of \SI{40}{\centi\meter\per\second} and for initial $\phi_{\ce{Fe2O3}}=1$.
The only difference between the simulations is that one simulation is performed right side up, i.e., gravity pointing downwards, and the second for the setup placed upside down, i.e., gravity pointing upwards.
From now on, we will refer to the right-side-up case as the gravity-pointing downwards case and the upside-down case as the gravity-pointing upwards case.
Similarly to our previous publication~\cite{2025_Hazenberg_AnalyzingIronDust}, the simulation is solved transiently until a quasi-steady state flame is achieved.
When the is at its quasi-steady position, a step increase of \num{30}\% to the inlet particle loading is applied.
The step change is a homogenous modification of the inlet particle loading, i.e., no spatial variations are introduced.
After this step change, the flame response is recorded for \SI{400}{\milli\second}.
The results presented in this article are circumferentially averaged in all our graphs to minimize statistical fluctuations inherent to these dispersed flames.
In this section, we will first analyze how a sudden change in particle loading modifies the position of the flame front. 
Second, we present how such change modifies the flame temperature. 
Thirdly, we show how the burning velocity reacts to this change.

\subsection{Flame position response}
\begin{figure}
    \centering
    \includegraphics{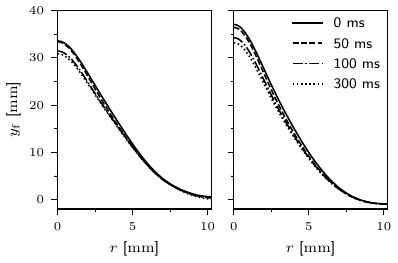}
    \caption{Evolution of the \SI{400}{\kelvin} iso-contour as the flame responds to an abrupt change in particle loading. 
    In the left frame, the gravity is pointing down, and in the right frame gravity is pointing up.}
    \label{Fig:Flame_position}
\end{figure}

Fig.~\ref{Fig:Flame_position} shows the \SI{400}{\kelvin} temperature iso-contour for four instances after the abrupt change in particle loading.
The left frame shows the case where gravity is pointing down, and the right frame shows the case where gravity is pointing up.
At $t=\SI{0}{\milli\second}$, the shapes of the temperature iso-contour are slightly different between the two cases.
More specifically, the height of the flame for the gravity downwards case is less than that for the gravity upwards case.
One can explain this from the residence time of particles inside the flame:
When gravity points downwards, the iron particles move slower than the gas, thus increasing the residence time inside the flame.
As a result, particles have more time to heat up, and when they start reacting, they move less, increasing the exchange term's magnitude.
This increase in the magnitude of the exchange term enhances the mass-burning rate.
In other words, the burning velocity for the gravity pointing down case is higher than for the gravity pointing up case.
One can also see this from Fig.~\ref{Fig:Flame_position}; the Bunsen flame is smaller.
An interesting thing to note is that the stand-off distance from the burner rim, located at $r=\SI{10.25}{\milli\meter}$ and $y=\SI{0}{\milli\meter}$, is smaller for the gravity-up case.
The smaller stand-off distance suggests that the attachment to the burner rim is stronger for the upside-down case than for the right-side-up case, even though the burning velocity is lower.
A detailed analysis of the flame stabilization mechanism at the burner rim is outside the scope of this article.

After the step change in particle loading, the flame requires some time to respond, as the additional particles need to be convected to the flame front.
Assuming that the particles experience no slip with respect to the gas, this delay is roughly \SI{42}{\milli\second} in the center of the domain.
As a result, the flame barely moves for the first \SI{50}{\milli\second} after the sudden change.
Most of the flame movement occurs between \num{50} and \SI{100}{\milli\second}, where the tip of the flame moves roughly \SI{2}{\milli\meter} closer to the burner.
After \SI{100}{\milli\second}, the flame is still moving but only with an additional millimeter in \SI{200}{\milli\second}.
The fact that the flame becomes smaller is easily explained; the increase in the particle loading enhances the burning velocity.
The flame moves only a small amount because the burning velocity is relatively insensitive to the particle loading, see~\cite{2023_Fedoryk_ExperimentalInvestigationLaminar,2025_Hazenberg_AnalyzingIronDust}.
We should also note that during the flame's movement towards its new quasi-steady position, the shape of the flame is maintained, and no wrinkling occurs.

\subsection{Temperature and oxygen response} \addvspace{10pt}
From Fig.~\ref{Fig:Flame_position}, we observed that the position of the flame front adjusts requires at least \SI{300}{\milli\second} to achieve a new quasi-steady position.
The position of the flame can only change if the burning velocity is modified such that the flame consumes gas either faster or slower than the local gas velocity.
In gas-phase flames, the flame temperature severely influences the burning velocity.
While it is known that the iron flames are not very sensitive to the flame temperature, we will still first analyze the response of the flame temperature to the particle loading.
Our primary interest is understanding how the flame adapts to the new particle loading.
We will first present how the flame temperature adapts as a function of the radial positions and thereafter study the temporal behavior.

\begin{figure}
    \centering
    \includegraphics{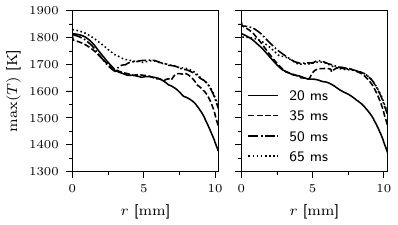}
    \caption{The flame temperature as a function of radius for four different instances of time.
    In the left frame, the gravity is pointing down, and in the right frame the gravity is pointing up.}
    \label{Fig:Max_Temperature_vs_pos}
\end{figure}
In Fig.~\ref{Fig:Max_Temperature_vs_pos}, the flame temperature as a function of the radial coordinate is shown at four different time instances.
The flame temperature is defined as the maximum temperature at a radial position, and the time is defined identically in Fig.~\ref{Fig:Flame_position}.
At \SI{20}{\milli\second}, the increased particle loading has not yet reached the flame front.
The flame temperature shows a profile where it is highest in the center ($r=\SI{0}{\milli\meter}$) - as this is the most isolated from the cold co-flow - and cools down towards the burner edge, severely impacted by heat loss to the burner rim.
From the flame temperature at \num{35} and \SI{50}{\milli\second}, it is visible that the temperature of the Bunsen flame is affected by the higher particle loading arriving at the flame front.
The additional particles arrive first at the burner rim ($r=\SI{10.25}{\milli\meter}$), where the temperature increases from $\pm\num{1350}$ to $\pm\SI{1500}{\kelvin}$.
As time progresses, the flame front's temperature increases from the burner rim toward the flame center.
No significant differences are observed in the flame response between the gravity down and gravity up case.
Comparing the increase in temperature at the edge to that in the center, one can note that the temperature increases significantly more near the burner rim.
Finally, when comparing the temperature for $r>\SI{5}{\milli\meter}$ at $t=\num{50}$ to \SI{65}{\milli\second}, we can see that the temperature remains mostly stable after achieving its new equilibrium value.

\begin{figure}
    \centering
    \includegraphics{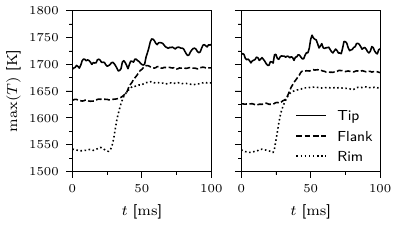}
    \caption{The average flame temperature as in function of time in three different areas of the flame:
    1) The tip ($0 < r < 0.33 R_{\mathrm{max}}$), 2) the flank ($0.33 R_{\mathrm{max}} < r < 0.66 R_{\mathrm{max}}$), and 2) the rim ($0.66 R_{\mathrm{max}} < r < R_{\mathrm{max}}$).
    In the left frame, the gravity is pointing down, and in the right frame gravity is pointing up.}
    \label{Fig:Max_Temperature_vs_time}
\end{figure}
To obtain more insight into the temporal behavior of the flame temperature, a graph of temperature as a function of time is presented in Fig.~\ref{Fig:Max_Temperature_vs_time}.
The graph is obtained by averaging the flame temperature in three separate parts of the flame:
1) The tip of the flame, here defined as $r < 0.33 R_{\mathrm{max}}$, 2) the flank of the flame ($0.33 R_{\mathrm{max}} < r < 0.66 R_{\mathrm{max}}$), and 3) the rim of the flame ($r > 0.66 R_{\mathrm{max}}$).
This graph confirms several of our earlier observations: The flame heats up the earliest and most at the burner rim and the last and least in the center.
Moreover, the temperature adjusts itself rapidly after the additional particles arrive.
Another thing to notice is that the flame temperature close to the center of the flame is somewhat unstable, while at the other positions, it is relatively stable.
The fluctuations in flame temperature at the center are not the result of, e.g., thermoacoustic instabilities triggered by the sudden change in equivalence ratio; they are due to inherent fluctuations in iron dust flames.
Even though ideal seeding is assumed, the position and size of a newly added particle are chosen randomly during the simulation.
As a result, some fluctuations are present in particle loading and average particle size at any position, as not enough particles are present only to observe the statistically averaged behavior.
These fluctuations are the reason why all the presented results are circumferentially averaged.
The circumferential averaging reduces the noise for increasing radius as the number of particles found between $r$ and $r+\delta r$ scales by $r$ itself, meaning for small $r$, the statistical fluctuations increase in the presented results.
Note that these fluctuations are not the same as those observed the discrete regime, e.g.,~\cite{2009_Tang_Effectdiscretenessheterogeneous}, as we are not resolving the particle boundary layers and can thus not observe discrete flame propagation.

From Fig.~\ref{Fig:Flame_position}, we concluded that the flame requires at least \SI{300}{\milli\second} to stabilize at the new quasi-steady position.
In contrast, in Fig.~\ref{Fig:Max_Temperature_vs_pos} and Fig.~\ref{Fig:Max_Temperature_vs_time}, we observe that the flame temperature adjusts itself within \SI{50}{\milli\second} after the increased particle loading arrives.
The rapid increase in flame temperature suggests that the burning velocity is rapidly increased, and thereafter, the flame steadily moves towards its new equilibrium position.
The lack of oscillations, except for the statistical ones in the center, in the flame temperature in Fig.~\ref{Fig:Max_Temperature_vs_time} supports this hypothesis.
That no significant oscillations are present in the flame temperature after a sudden change in particle loading is noteworthy, as conventional gas phase flames are expected to exhibit (strong) oscillations.

\begin{figure}
    \centering
    \includegraphics{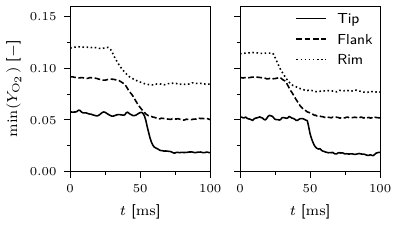}
    \caption{The average minimal \ce{O2} fraction as in function of time in three different areas of the flame:
    1) The tip ($0 < r < 0.33 R_{\mathrm{max}}$), 2) the flank ($0.33 R_{\mathrm{max}} < r < 0.66 R_{\mathrm{max}}$), and 2) the rim ($0.66 R_{\mathrm{max}} < r < R_{\mathrm{max}}$).
    In the left frame, the gravity is pointing down, and in the right frame gravity is pointing up.}
    \label{Fig:O2min_vs_time}
\end{figure}
To better understand how fast the mass burning rate of the flame adjusts to the new particle loading, the remaining or minimal fraction of \ce{O2} is analyzed. 
Similar to the flame temperature vs time graph (Fig.~\ref{Fig:Max_Temperature_vs_time}), the minimal fraction of \ce{O2} is evaluated at every radial position and then averaged in three different parts: the 1) tip, 2) flank and 3) rim of the flame.
In Fig.~\ref{Fig:O2min_vs_time}, the averaged minimal \ce{O2} mass fractions are shown as a function of time.
We can make similar observations as in our previous publication~\cite{2025_Hazenberg_AnalyzingIronDust}; the oxygen mass fraction increases towards the Bunsen cone's edge due to mixing of the co-flow air in the burned gas.
In line with the temporal evolution of the flame temperature (see Fig.~\ref{Fig:Max_Temperature_vs_time}), the oxygen mass fraction in the burned gas first reduces at the burner edge and then in the flame center.
After the increased particle loading arrives at the flame front ($\pm \SI{40}{\milli\second}$), the entire flame structure of the Bunsen flame adjusts itself in less than \SI{50}{\milli\second}.
Locally, this happens much faster; notice how the flame tip, where the additional particles arrive roughly simultaneously, adjusts within \SI{10}{\milli\second}, and no significant oscillations follow.
From the temporal evolution of temperature and oxygen, we conclude that the flame adjusts rapidly to the increased particle loading.
Moreover, the sudden change in particle loading does not appear to introduce any oscillations in the \ce{O2} mass fraction inside the burned gas.
The lack of an oscillatory response further confirms that the flame exhibits a strongly damped response to equivalence ratio fluctuations.
As such, the flame is expected to be robust to small perturbations on the inflow.

\subsection{Burning velocity response}
In this last part of the results section, we will discuss how the burning velocity of the flame evolves as a response to the sudden change in equivalence ratio.
The burning velocity is extracted using an extended method from our previous article~\cite{2021_Hazenberg_StructuresBurningVelocities}, now accounting for the flame movement.
At any radial position, the burning velocity is estimated via
\begin{equation}
    S_{\mathrm{L},\theta} = \left(v_{\mathrm{in}}(r) - v_{\mathrm{f}}(r) \right) \cos \left(\theta(r)\right) \mathrm{,}
    \label{Eq:Flame_speed_angle_based}
\end{equation}
where $v_{\mathrm{in}}(r)$ is the inlet velocity, $v_{\mathrm{f}}(r)$ is the velocity parallel to the inlet velocity of the chosen temperature iso-contour, and $\theta(r)$ is the local flame angle with respect to the flow.
To get the global burning velocity of the Bunsen flame, Eq.~\ref{Eq:Flame_speed_angle_based} is averaged between $r=0.33R_{\mathrm{max}}$ and $r=0.66R_{\mathrm{max}}$.
The goal of obtaining the burning velocity like this is not to get an accurate transient value of the burning velocity but mainly to gain insight into the time it takes for the flame to stabilize in its new position.

\begin{figure}
    \centering
    \includegraphics{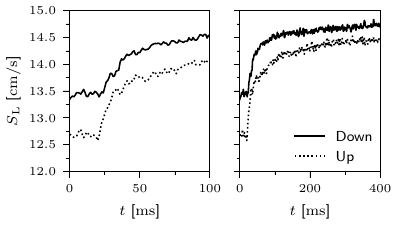}
    \caption{Evolution of the global burning of the bunsen flame for both the gravity up and down case.
    In the left frame a zoom is provided of the first \SI{100}{\milli\second} and the right frame shows the entire \SI{400}{\milli\second}.}
    \label{Fig:Burning_velocity_vs_time}
\end{figure}
In Fig.~\ref{Fig:Burning_velocity_vs_time}, the global burning velocity as a function of time is shown.
These graphs confirm our earlier observation that the burning velocity is lower for the gravity-pointing-up case.
The flame's burning velocity adjusts to the new particle loading directly after the additional particles arrive at the flame front, roughly \SI{25}{\milli\second}.
Initially, the increase in burning velocity is swift (between \num{25} and \SI{75}{\milli\second}), corresponding to the fast adaptation of the flame front to the increased particle loading, also observed in Figs.~\ref{Fig:Max_Temperature_vs_time}~and~\ref{Fig:O2min_vs_time}.
After \SI{75}{\milli\second}, the rate at which the burning velocity increases has reduced but does keep increasing until \SI{400}{\milli\second}.
We believe that the initial increase ($25 < t < \SI{75}{\milli\second}$) is due to the modification of the burning velocity by the additional particles, and the second increase ($75 < t < \SI{400}{\milli\second}$) is due to changing 3D effects as the flame searches for its new equilibrium position.
With 3D effects, we mean that stretch and curvature evolve as the shape of the flame fronts adapts to its new equilibrium position. 
Similar to our previous observations, the burning velocity shows no oscillatory behavior except for the statistical fluctuation mentioned before. 
Again the lack of oscillations further supports our general observation that these iron dust flames are surprisingly robust to sudden changes in particle loading.
Finally, we have not identified any significant differences in flame response between the right-side-up and upside-down configurations.

\section{Discussion and Conclusion} \addvspace{10pt}
In this article, we have investigated the sensitivity of an iron dust Bunsen flame to sudden changes in particle loading.
Due to the well-known challenges in obtaining 1) a stable particle loading and 2) stabilizing these flames experimentally, we expected a strong flame response to sudden changes in particle loading.
Our idea was that small fluctuations in the particle loading would be amplified by the flame, resulting in a highly unstable flame.
In other words, we expected the flame to oscillate, similar to gas phase flames, after a sudden change in particle loading/fuel equivalence ratio.
However, our results show that the flame does not oscillate and exhibits a damped response toward its new equilibrium position.
We hypothesize that this damped response is due to the insensitivity of these flames to the burned gas temperature:
In a gas phase flame, an increase in the burned gas temperature results in an exponential increase in the mass burning rate.
However, oxygen diffusion to the particles controls the mass burning rate in these flames.
The diffusion-controlled mass burning rate means that the mass burning rate roughly relates to temperature via $\dot{m}_{\mathrm{b}} \propto \rho D \propto T_{\mathrm{b}}^{0.7}$, i.e., only weakly depends on temperature.
On top of that, the increased particle loading reduces the \ce{O2} mass fraction in the burned gas, reducing the mass burning rate of a single particle.
As a result, the conventional thermo-acoustic feedback mechanism of a chemistry-driven gas-phase flame might be absent in iron dust flames.
Further studies are required to confirm this hypothesis, e.g., determining the flame transfer function in a straightforward 1D domain.

Our observation raises new questions about the mechanism by which experimental flames are unstable.
From the simulations presented in this article, we conclude that the flame position response is damped when sudden changes in particle loading are applied.
Additionally, based on the observations from Fedoryk et al.~\cite{2023_Fedoryk_ExperimentalInvestigationLaminar}, we know that spatial fluctuations are also present.
Based on the results presented in this work, we believe that the observed instabilities in the experimental flames might exclusively be due to the seeding and flow field, i.e., the particle loading and flow in front of the flame are unstable.
We removed such instabilities by limiting ourselves to homogenous changes in particle loading.
Future research can study the impact of spatial fluctuations either by artificially generating such fluctuations or by including the combustion pipe (see Fig.~\ref{fig:expSetup}) in the simulation.

Finally, we performed this study for the right-side-up (gravity down) and upside-down (gravity up) configuration.
From this, we noted that the burning velocity is reduced when the flow is in the direction of the gravity due to a reduced residence time of particles in the flame front.
However, we did not identify any significant differences in flame stability for the conditions under consideration in this article.

\acknowledgement{Declaration of competing interest}
The authors declare that they have no known competing financial interests or personal relationships that could have appeared to influence the work reported in this paper.

\acknowledgement{Acknowledgments}
This work was funded by the Hessian Ministry of Higher Education, Research, Science and the Arts - cluster project Clean Circles.

\section{CReDiT authorship contribution statement}
\textbf{T. Hazenberg:} Conceptualization, Data curation, Formal analysis, Investigation, Methodology, Software, Validation, Visualization, Writing - original draft.
\textbf{D. Braig:} Conceptualization, Data curation, Formal analysis, Investigation, Methodology, Software, Validation, Writing - reviewing \& editing.
\textbf{J. Mich:} Software, Validation, Writing - reviewing \& editing.
\textbf{A. Scholtissek:} Conceptualization, Funding acquisition, Project administration, Supervision, Writing - reviewing \& editing.
\textbf{C. Hasse:} Conceptualization, Funding acquisition, Project administration, Resources, Supervision, Writing - reviewing \& editing.

\footnotesize
\baselineskip 9pt

\bibliographystyle{elsarticle-num}
\bibliography{references}

\newpage

\small
\baselineskip 10pt

\end{document}